\documentclass{emulateapj}

\shorttitle{The distance and mass of PSR J1023+0038}

\newcommand{\Msun}{\ensuremath{M_{\odot}}}

\begin{document}

\submitted{ApJL, accepted 23 July 2012}
\title{A parallax distance and mass estimate for the transitional millisecond pulsar system J1023+0038}

\author{A.T. Deller\altaffilmark{1}, A.M. Archibald\altaffilmark{2}, W.F. Brisken\altaffilmark{3}, S. Chatterjee\altaffilmark{4}, G.H. Janssen\altaffilmark{5}, V.M. Kaspi\altaffilmark{2}, D. Lorimer\altaffilmark{6}, A.G. Lyne\altaffilmark{5}, M.A. McLaughlin\altaffilmark{6}, S. Ransom\altaffilmark{7}, I.H. Stairs\altaffilmark{8}, B. Stappers\altaffilmark{5}}
\altaffiltext{1}{The Netherlands Institute for Radio Astronomy (ASTRON), Dwingeloo, The Netherlands}
\altaffiltext{2}{Department of Physics, McGill University, 3600 University Street, Montreal, QC H3A 2T8, Canada}
\altaffiltext{3}{National Radio Astronomy Observatory, Socorro, NM 87801, USA}
\altaffiltext{4}{Astronomy Dept., Cornell University, Ithaca, NY 14853, USA}
\altaffiltext{5}{Jodrell Bank Centre for Astrophysics, School of Physics and Astronomy, University of Manchester, Manchester M13 9PL, UK}
\altaffiltext{6}{Department of Physics, West Virginia University, Morgantown, WV 26506, USA}
\altaffiltext{7}{National Radio Astronomy Observatory, Charlottesville, VA 22903, USA}
\altaffiltext{8}{Department of Physics and Astronomy, University of British Columbia, 6224 Agricultural Road, Vancouver, BC V6T 1Z1, Canada}

\begin{abstract}
The recently discovered transitional millisecond pulsar system J1023+0038 exposes a crucial evolutionary phase of recycled neutron stars for multiwavelength study.  The system, comprising the neutron star itself, its stellar companion, and the surrounding medium, is visible across the electromagnetic spectrum from the radio to X--ray/gamma--ray regimes and offers insight into the recycling phase of millisecond pulsar evolution. 
Here, we report on multiple--epoch astrometric observations with the Very Long Baseline Array (VLBA) which give a system parallax of $0.731 \pm 0.022$ milliarcseconds (mas) and a proper motion of $17.98\pm0.05$ mas yr$^{-1}$. By combining our results with previous optical observations, we are able to use the parallax distance of $1368^{+42}_{-39}$~pc to estimate the mass of the pulsar as $1.71 \pm 0.16~\Msun$, and we are also able to measure the 3D space velocity of the system as $126 \pm 5$ km s$^{-1}$.  
Despite the precise nature of the VLBA measurements, the remaining $\sim$3\% distance uncertainty dominates the 0.16~$\Msun$ error on our mass estimate.
\end{abstract}


\section{Introduction}

``Recycled" millisecond pulsars attain their high spin frequencies by accreting matter from a companion donor star \citep{alpar82a,bhattacharya91a}.  The mass--transfer phase is identified by variable X--ray emission from the heated gas in the pulsar accretion disk, and this phase is thought to be relatively well characterized theoretically.  However, the end of accretion and birth of the radio--visible millisecond pulsar is still poorly understood.  The discovery of a millisecond pulsar system whose optical variability suggests it had an accretion disk in 2001
\citep[PSR J1023+0038\footnote{The system is also known by its FIRST designation, FIRST J102347.6+003841; here, we use J1023 when referring to the pulsar and ``the J1023 system" when referring to the binary system}, hereafter J1023;][]{archibald09a, wang09a} offers a unique opportunity to study this short-lived transitional regime.

The J1023 system comprises a 1.69--ms radio pulsar with a nondegenerate companion in a 4.75--hour orbit.   It displays unusual observational characteristics which are consistent with its apparent evolutionary stage -- the pulsar dispersion measure varies on timescales of minutes to weeks, and long--term variation in the orbital period is also apparent. Eclipses of the pulsar emission are observed, usually at particular orbital phases \citep{archibald09a}.  The influence of material liberated from the companion object and interacting with the pulsar wind is likely responsible for all of these phenomena.  The J1023 system, then, offers a rich source of information for the study of binary evolution and interacting wind physics.

As a compact radio source of $\sim$mJy brightness, J1023 is a suitable target for astrometric observations using very long baseline interferometry (VLBI), which can provide extremely high distance accuracy via the measurement of annual geometric parallax.
Pulsar astrometric campaigns with the Very Long Baseline Array (VLBA) in the US \citep{chatterjee09a,brisken02a} and the Long Baseline Array (LBA) in Australia \citep{deller09b} have obtained parallax measurements with accuracies at the level of 20~$\mu$as.

\section{Observations and data processing}

\subsection{VLBI observations}

Due to the very high resolution of VLBI observations ($\le$10 mas), even a detection of moderate significance yields a very precise position centroid, and systematic errors relating to calibration typically dominate the astrometric error budget \citep{chatterjee04a}.  By far, the dominant contribution to systematic calibration errors at frequencies below several GHz is the unmodeled differential ionosphere between the calibrator and target sources.  This contribution can be reduced by observing at higher frequencies or utilizing a calibrator very close to the target on the sky.
As Table~\ref{tab:observations} illustrates, J1023 has an average (although highly variable) flux of several mJy at 1.6 GHz.  Our 1.6 GHz astrometric observations detected J1023 with a significance ranging from 7 to 63, and since J1023 possesses a typical steep pulsar spectrum, higher frequency observations would not have regularly detected the pulsar.  Accordingly, minimizing the ionospheric error was only possible by minimizing the calibrator--target separation.

To this end, we made use of an ``in--beam" calibrator (whose angular separation from the target is less than the size of the telescope primary beam, and can thus be observed simultaneously with the target).  
By analyzing archival Very Large Array observations we identified the flat spectrum object PMN J1023+0024 as a likely compact source, and confirmed its suitability with exploratory VLBA observations.  PMN J1023+0024 is bright (peak flux density 80 mJy/beam at 1.6 GHz) and almost unresolved (deconvolved size 2.5$\times$1.5 mas).  It is separated from J1023 by just 15\arcmin\ (J2000 position 10:23:38.77064 +00:24:16.6690; estimated uncertainty 0.5 mas in each coordinate).

\begin{deluxetable}{llll}
\tabletypesize{\small}
\tablecaption{Observation details}
\tablewidth{0pt}
\tablehead{
\colhead{Observation} & \colhead{Observation} & \colhead{MJD} & \colhead{Pulsar flux} \\
\colhead{group} & \colhead{dates} &  & \colhead{(mJy)\tablenotemark{a}}
}
\startdata
2008a	&  Dec. 16			&	54816 			& 2.0 \\
2009a	&  Feb. 14			&	54876 			& 0.7 \\
2009b	&  May 17, Jun. 12	&	54968, 54995 		& 3.0, 1.1 \\
2009c	&  Nov. 21, Dec. 7	&	55156, 55172 		& 0.9, 1.9\\
2010a	&  May 15, 29		&	54816, 55346 		& 0.4, 3.1\\
2010b	&  Nov. 17, Dec. 3	&	55517, 55533		& 2.5, 1.1
\enddata
\tablenotetext{a}{\small{Averaged over the pulsar period}}
\label{tab:observations}
\end{deluxetable}

We conducted 10 astrometric observations of J1023 spread over a 2--year period from late 2008 to late 2010.  In each observation, both right and left circular polarization were recorded in four observing bands of width 8 MHz, resulting in a total data rate of 256 Mbps/antenna.  The bands were placed adjacent to one another and spanned the frequency range  1651.49 -- 1683.49 MHz.  After two exploratory observations, the final eight observations were clustered in four pairs, with each pair being observed at a time close to parallax extrema to improve our sensitivity to the parallax signature.  The observations are summarized in Table~\ref{tab:observations}.  The observations were phase referenced to the bright VLBA calibrator J1024--0052, and the pointing center for the target source was placed midway between J1023 and PMN J1023+0024.  Since the in--beam calibrator was relatively bright, an unusually long cycle time (10 minutes) back to the primary calibrator was used, as this ``phase referencing" was used solely to improve the absolute position of the in--beam calibrator and to refine the amplitude calibration.  A total of 180 minutes on--source time was obtained per epoch -- after the excision of data during the pulsar eclipse (discussed below), the resultant continuum image sensitivity was typically 75 $\mu$Jy, although a number of epochs suffered instrument failure at one or more antennas, slightly reducing sensitivity.

\subsection{Supporting timing observations}

In contrast to the very stable timing behavior exhibited by most millisecond pulsars, J1023 exhibits small, apparently random fluctuations in its orbital period on short (week to month) timescales, making extrapolation of timing solutions difficult \citep{archibald09a}.  
In order to predict the pulsar rotational phase accurately and employ pulsar gating to boost S/N \citep[see e.g.,][]{deller09b}, we carried out regular timing observations of J1023 in support of our VLBA program.  These observations are briefly summarized below.
 
For the first VLBA epochs we used the timing solution published in \citet{archibald09a}. For later epochs, we observed J1023 for an hour every few weeks with the Arecibo telescope and the ASP coherent dedispersion backend \citep{demorest07a,ferdman08a} at $1.4~\mathrm{GHz}$. This yielded adequate ephemerides until early 2010, when damage to the Arecibo telescope was discovered, requiring zenith angle restrictions that precluded observations of J1023. 
J1023's erratic dispersion measure variations made phase connection to ongoing 350 MHz Green Bank Telescope observations unreliable, and so for the final epochs we made use of data from the J1023 timing programs carried out using the Lovell Telescope at Jodrell Bank \citep{hobbs04a} and the Westerbork Synthesis Radio Telescope \citep[WSRT;][]{karuppusamy08a}. 

An additional complication is that J1023 is undetectable at $1.4\;\mathrm{GHz}$ for about 30\% of the orbit \citep{archibald09a}. After correlation, we discarded all integrations that fell between orbital phases $0.15$ and $0.45$ (around 60 minutes of on--target time per observation) using a simple orbital ephemeris that assumes a constant orbital period \citep{archibald10a}.

\subsection{Correlation and data processing}

From the first three observation groups, all epochs except 2009 May 17 were correlated using the VLBA hardware correlator.  The epoch at 2009 May 17 and the final 6 epochs were correlated using the DiFX software correlator \citep{deller11a}.  In each case, two correlator passes were employed; the first without pulsar gating at the position of PMN J1023+0024, and the second using the pulsar gating scheme described below at the position of J1023.  An averaging time of 2 seconds was used for both passes.

Pulsar gating involves zeroing the correlator output when the pulse phase is outside a specified window, preventing the addition of unnecessary noise into the visibility outputs.  J1023 has a wide, double peaked pulse shape \citep{archibald09a}, which limits the gain from pulse gating to a factor of $\sim$1.7 when a simple on/off gate is used.  The erratic timing behavior of J1023 precludes the use of more advanced ``matched filtering" \citep{deller07a}, which could otherwise have boosted the gate gain to a factor of $\sim$2.2.

The visibility data produced by the correlator were reduced using AIPS\footnote{http://www.aips.nrao.edu/}, utilizing standard scripts based on the ParselTongue package \citep{kettenis06a}. After loading the data and flagging known bad data (including time ranges affected by the eclipse), the visibility amplitudes were calibrated, fringe fitting and bandpass calibration were performed using the bright source J0927+3902, and the delay and amplitude solutions were further refined with fringe fitting and amplitude self--calibration using the out--of--beam reference source J1024--0052.  Subsequently, phase--only corrections were generated from the in--beam calibrator PMN J1023+0024. This and all future steps were undertaken on the Stokes I data.  For all reference sources, a combined model was formed based on the data from all epochs; these models were not permitted to vary between epochs.

Finally, the J1023 data were split and averaged in frequency, leaving a single visibility point per band for each integration, and the visibility data were written to disk and imaged using difmap \citep{shepherd97a} with natural weighting (which maximizes the image sensitivity).  The resultant images (one per 8--MHz band, Stokes I, hereafter the ``single--band" images) were fitted with a single gaussian component using the AIPS task JMFIT, as was a combined image formed using all subbands (hereafter the ``combined" image).  For two epochs (MJD 54876 and 55331) the pulsar was not detected in the single--band images, and only the combined image was formed and fitted for position.

\subsection{Astrometric data reduction}

The astrometric error in the position measurement for each epoch is principally composed of a random component (the error in fitting a position centroid in the image plane) and a systematic component (from the unmodeled delay errors at each telescope due to the atmosphere and ionosphere).  The pulsar brightness varies by an order of magnitude throughout our 10 observing epochs; random errors dominate the error budget when it is faint, whilst for the remaining epochs the systematic error dominates.  This fact, along with the highly variable atmospheric and ionospheric conditions between epochs, means that the distribution of astrometric errors in right ascension and declination will not be Gaussian.

A straightforward fit to the 10 positions obtained from the combined images yields a reduced $\chi^{2}$ value of 2.2, indicating that the errors on a whole are underestimated due to the unaccounted--for systematic contribution.  Simply inflating the error budget by scaling the position errors at each epoch to obtain a reduced $\chi^{2}$ value of 1.0 is inappropriate, since that assumes the systematic error contribution is correlated with the random error contribution at each epoch.  A better, although still imperfect approach is to add an equal error in quadrature to all epochs.  In order to obtain a reduced $\chi^{2}$ value of 1.0, the required error magnitude was 50 $\mu$as, added solely to the right ascension coordinate\footnote{The fit varies insignificantly if a larger total error of 70 $\mu$as is divided equally between right ascension and declination.}.  This astrometric error corresponds to an average residual path length error of 2 mm.
At our observing frequency of 1.6 GHz, the ionospheric correction will be by far the dominant source of path length error, as current ionospheric models sample the sky coarsely in time and area. During our observations, the differential ionospheric delay correction between the calibrator and the target ranged from 0 to 80 picoseconds (0 to 25 mm), with typical values of $\sim$20 picoseconds (6 mm).  A path length error of 2 mm, then, corresponds to an average error of around 30\% in the ionospheric correction.  The fitted parameters obtained when a constant systematic error contribution of 50 $\mu$as is added to each epoch (fit ``A") are shown in the first data column of Table~\ref{tab:fits}.  The position errors shown in parentheses for right ascension and declination are 
formal errors shown for comparative purposes only.  The actual errors in the absolute position coordinates are dominated by the absolute position of the reference source PMN J1023+0038 and are estimated at 0.5 mas.

An alternate method to estimating the astrometric errors is to use bootstrapping to estimate the actual form of the error distribution without imposing any prior constraints.  Bootstrapping is described in \citet{efron91a} and has been previously used in the estimation of errors in pulsar astrometry observables \citep[e.g.,][]{chatterjee09a,deller09b}.  In essence, it involves creating a large number of test datasets, where each dataset is constructed by sampling with replacement from the pool of measured astrometric positions.  The astrometric observables are fitted once from each test dataset and a large sample of tests is used to build a histogram of the fitted values for each observable.  The results of a bootstrap error analysis for J1023 using positions from the single--band  images (fit ``B") are shown in the second data column of Table~\ref{tab:fits}.

\begin{figure}
\begin{center}
\begin{tabular}{c}
\includegraphics[width=0.45\textwidth]{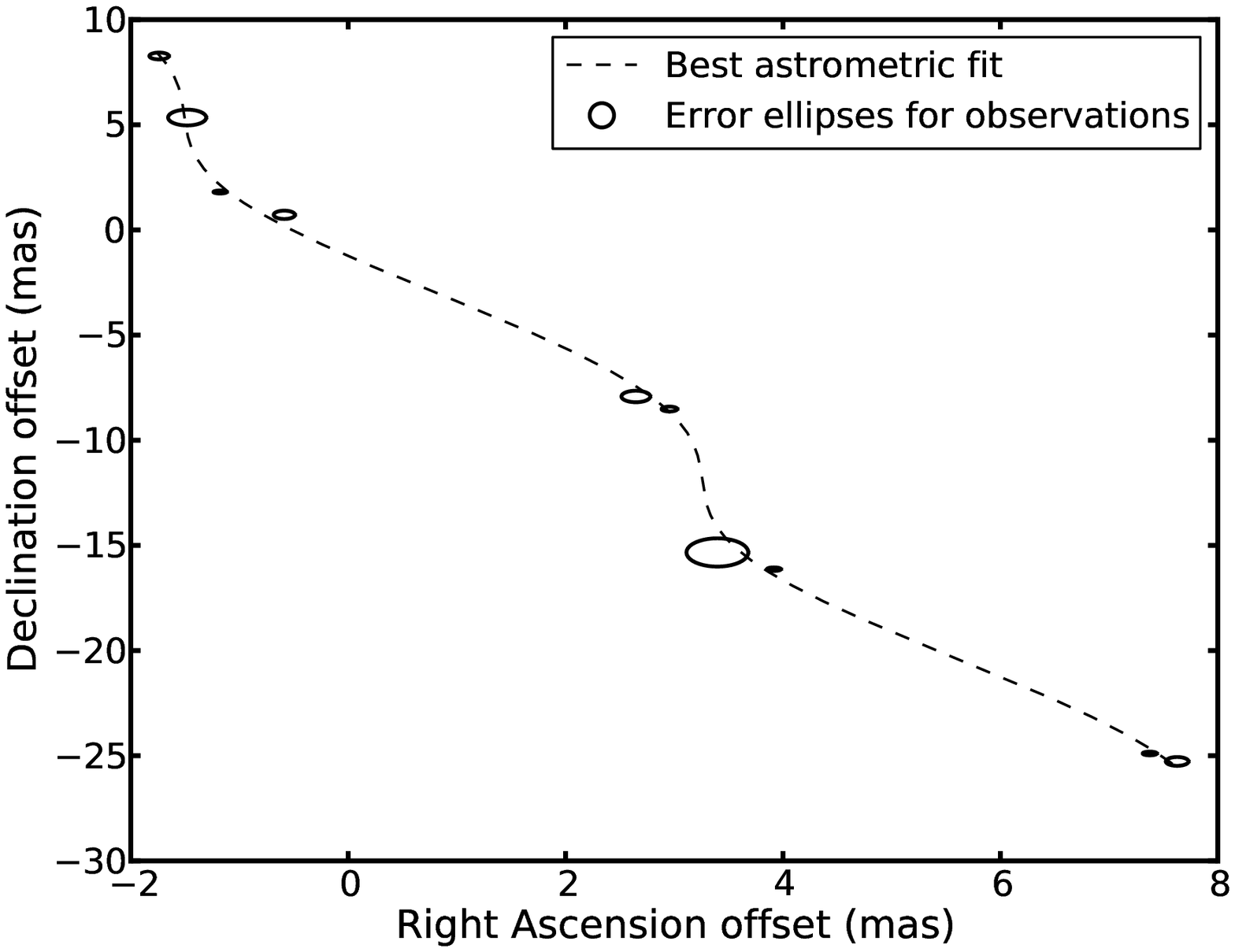} \\
\includegraphics[width=0.45\textwidth]{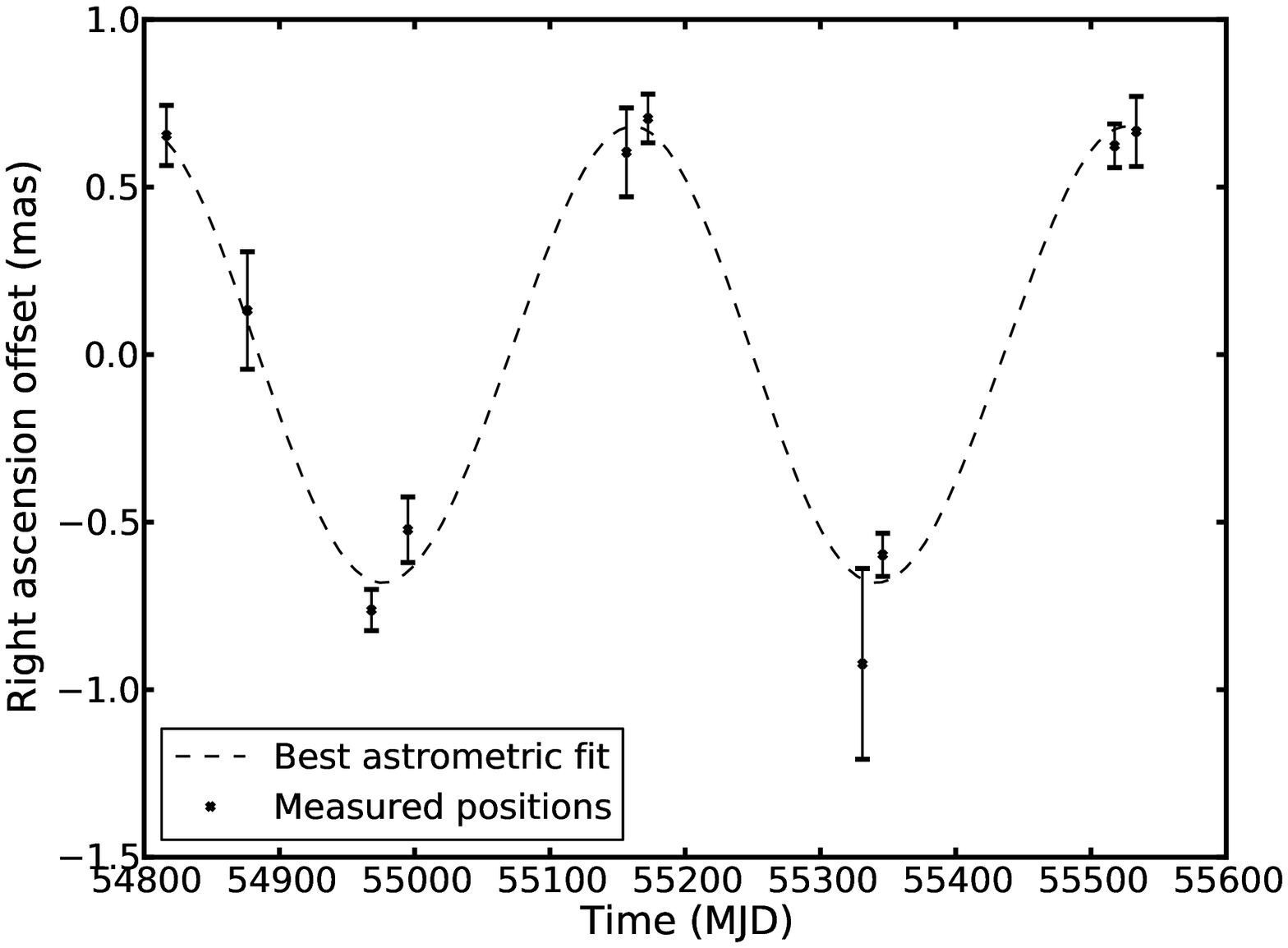} \\
\includegraphics[width=0.45\textwidth]{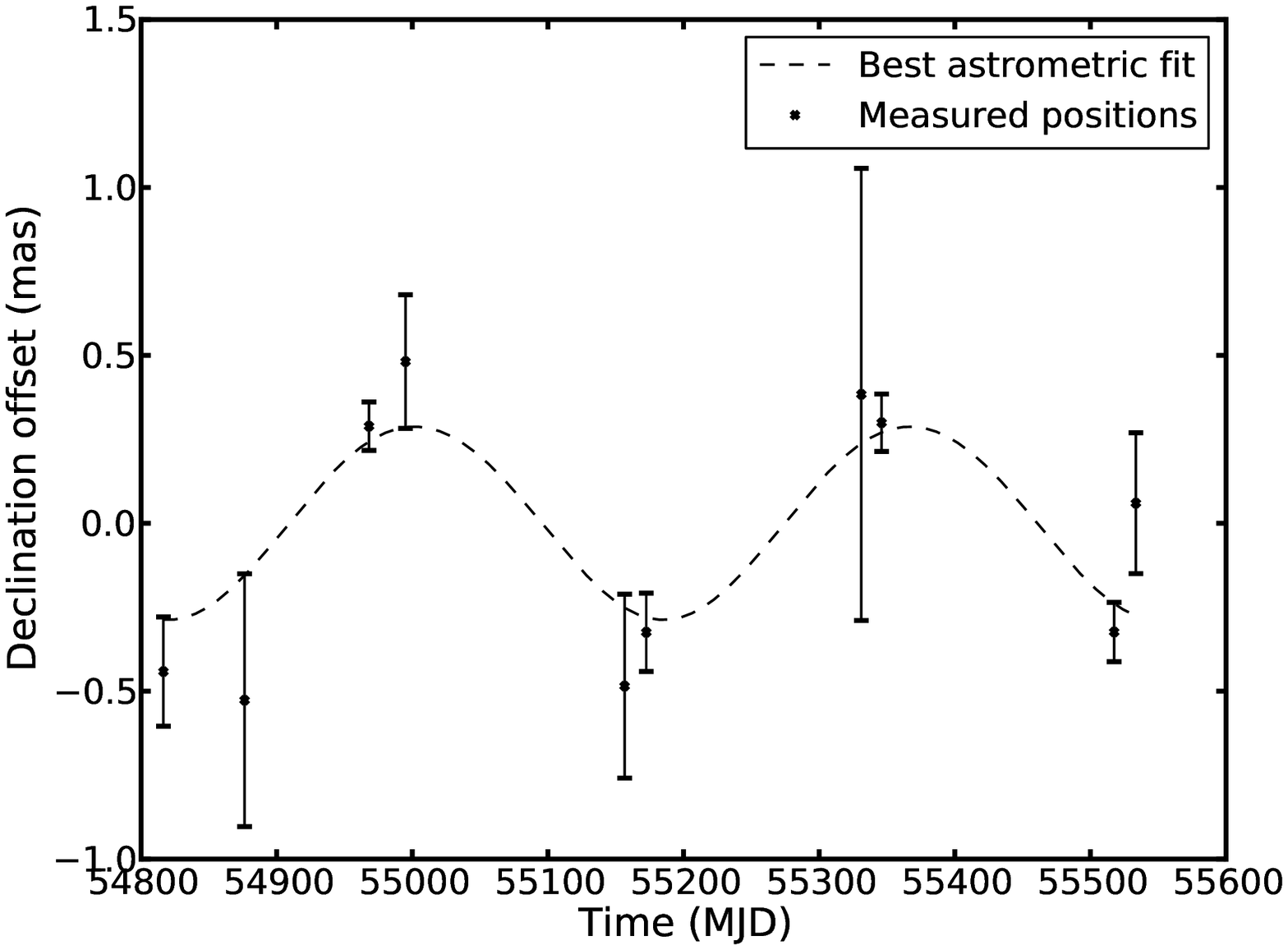} 
\end{tabular}
\end{center}
\caption{The astrometric fits to the J1023 VLBI data.  Top panel: motion relative to the reference position of 10:23:47.687198+00:38:40.84551 over the two years of observations, with the best fit shown by the dashed line.  Middle panel: The parallax signature in right ascension (proper motion subtracted). Bottom panel: The parallax signature in declination (proper motion subtracted).}
\label{fig:fit}
\end{figure}

The disadvantage of the bootstrapping technique is that a relatively large number of observed positions are required to avoid small--sample--size uncertainty.  This necessitates using the single--band position fits for the epochs where they are available, but the relatively low significance of the detections becomes problematic.  The positions (and associated errors) obtained from image plane fits become increasingly unreliable at low S/N ratios; in fact, the weighted mean position obtained by averaging the single--band positions can differ from the combined image position fit by up to 0.5$\sigma$.  Accordingly, we would expect the bootstrap error values to be more conservative than those in fit A, and this is indeed seen, although the difference is small and the fitted values are consistent.  Hereafter, we use the values and errors from fit A in all further analysis.  The motion of J1023 over 2 years and the parallax signature after the subtraction of the best--fit proper motion are plotted in Figure~\ref{fig:fit}.

\begin{deluxetable}{lrrr}
\tabletypesize{\small}
\tablecaption{Fitted astrometric parameters for J1023.}
\tablewidth{0pt}
\tablehead{
\colhead{Parameter} & \colhead{Fit A (combined)\tablenotemark{a}} & \colhead{Fit B (bootstrap)}
}
\startdata
Right ascension (J2000)\tablenotemark{b}			& 10:23:47.687198(2)	
							& 10:23:47.687196(2)		\\
Declination (J2000)\tablenotemark{b}				& 00:38:40.84551(4)	
							& 00:38:40.84551(4) 	\\
Position epoch (MJD)			& 55000 & 55000 \\
$\mu_{\alpha}$	(mas yr$^{-1}$)		& 4.76 $\pm$ 0.03
							& 4.79 $\pm$ 0.04   \\
$\mu_{\delta}$	(mas yr$^{-1}$)		& $-$17.34 $\pm$ 0.04	
							& $-$17.35 $\pm$ 0.05   \\
Parallax (mas)	 				& 0.731 $\pm$ 0.022	
							& 0.730 $\pm$ 0.028  \\
Distance (pc)					& 1368 $^{+42}_{-39}$
							& 1370 $^{+56}_{-51}$ \\
$v_{\mathrm T}$ (km s$^{-1}$)		& 117$^{+4}_{-4}$
							& 117$^{+5}_{-5}$
\enddata
\tablenotetext{a}{\small{Values from the combined fit (fit A) are used in the analysis.}}
\tablenotetext{b}{\small{The errors quoted here are from the astrometric fit only and do not include the $\sim$0.5 mas position uncertainty transferred from the inbeam calibrator's absolute position.}}
\label{tab:fits}
\end{deluxetable}

\section{Implications of the astrometric results}

The measured parallax of 0.731 $\pm$ 0.022 $\mu$as yields a distance to the J1023 system of $1368^{+42}_{-39}$~pc.  The very high precision of the measurement means that the Lutz--Kelker bias in the distance \citep{verbiest10a} is negligible for this system. Combined with the total measured proper motion of $17.98\pm0.05$ mas yr$^{-1}$, the transverse velocity is calculated as 117 $\pm$ 4 km s$^{-1}$.  The distance is over twice that predicted by the NE2001 electron density distribution model \citep[600 pc;][]{cordes02a}.  Accordingly, the average electron density along the line of sight is lower than expected, just 0.010 cm$^{-3}$.  The systemic radial velocity for J1023 reported by \citet{thorstensen05a} is negligible ($1\pm3$ km s$^{-1}$), and so the 3D system velocity is equal to the transverse velocity calculated here.  Correction for peculiar solar motion and Galactic rotation using a flat rotation curve and the current IAU recommended rotation constants (R$_{0}$ = 8.5 kpc, $\Theta_{0}$ = 220\,km\,s$^{-1}$) alters this value slightly to $126 \pm 5$ km s$^{-1}$.

Having a precise distance and velocity makes it possible to calculate corrections to the observed pulsar spin period derivative $\dot{\mathrm{P}}_{\mathrm{app}}$ due to the Shklovskii effect \citep{shklovskii70a} and acceleration in the Galactic potential \citep[e.g.,][]{nice95a}.   After these corrections are subtracted to obtain the true spin--period derivative $\dot{\mathrm{P}}$, this can be used to obtain accurate values for quantities such as the spin--down luminosity $\dot{\mathrm{E}}$, the magnetic field at the neutron star surface $\mathrm{B}_{\mathrm{surf}}$, and the characteristic age $\tau_{\mathrm{c}}$.  Using the supporting timing observations made with the Lovell Telescope, we obtain $\dot{\mathrm{P}}_{\mathrm{app}} = \left(6.83 \pm 0.05\right) \times 10^{-21}$.  The contribution due to the Shklovskii effect is $\dot{\mathrm{P}}_{\mathrm{Shk}}/\mathrm{P} = \mu^{2}\mathrm{D}/\mathrm{c}$, where $\mu$ is the proper motion, D is the pulsar distance and c is the speed of light.  Substituting the distance and velocity derived above, and taking P as 1.6879874 ms \citep{archibald09a}, we obtain $\dot{\mathrm{P}}_{\mathrm{Shk}} = \left(1.81 \pm 0.04\right) \times 10^{-21}$.  
Finally, we calculate the net effect of acceleration in the Galactic gravitational potential on $\dot{\mathrm{P}}_{\mathrm{app}}$ to be $\left(-2.8 \pm 0.3\right)\times10^{-22}$, following \citet{nice95a} and taking the vertical acceleration component from \citet{holmberg04a}.  We used R$_{0}$ = 8.5 kpc and $\Theta_{0}$ =  220\,km\,s$^{-1}$, and assumed 10\% errors in these constants and the $K_z$ force law of \citet{holmberg04a}.
Thus, the intrinsic $\dot{\mathrm{P}}$ is $\left(5.30 \pm 0.07\right) \times 10^{-21}$.

This revised $\dot{\mathrm{P}}$ is considerably lower than the preliminary value presented in \citet{archibald09a}, with consequent effects for $\dot{\mathrm{E}}$, $\mathrm{B}_{\mathrm{surf}}$, and $\tau_{\mathrm{c}}$.  $\dot{\mathrm{E}}$ is reduced to $4.3\times10^{34}$ ergs s$^{-1}$, $\mathrm{B}_{\mathrm{surf}}$ is reduced to $9.6\times10^{7}$ G, and $\tau_{\mathrm{c}}$ is increased to $5.0\times10^{9}$ years.  The value for  $\mathrm{B}_{\mathrm{surf}}$ is noteworthy given that less than 10\% of recycled pulsars have $\mathrm{B}_{\mathrm{surf}} < 10^{8}$ G.

\subsection{The mass of J1023}
\label{sec:mass}
When \citet{thorstensen05a} carried out optical modelling of
the J1023 system, they noted that the mass of the companion could be
inferred from the distance, assuming that it fills its Roche lobe.
Essentially, the distance combined with the multiband photometry
determine the temperature and size of the companion, and the
assumption links this to the size of the Roche lobe and therefore the
mass of the companion $M_c$.  Specifically, \citet{thorstensen05a} fit models including Roche
geometry and companion temperature distributions to the observed
multiband light curves and find that the distance to the system (under the assumption of Roche--lobe--filling) is:

$$
\mathrm{D} = 2.20 \left( \frac{M_c}{M_\Sun} \right)^{1/3} \mathrm{kpc}.
$$

Table 5 of \citet{thorstensen05a} lists 49 sets of model parameters
that fit the observed optical data. Testing the above relationship on
all 49 reveals a scatter of only 1\% on the coefficient (2.20\,kpc) over the broad range of parameters they consider
plausible enough to fit.

The pulsar timing measurements of \citet{archibald09a} showed that the
system mass ratio $M_{\mathrm{PSR}}/M_c = 7.1\pm 0.1$. Combining these
two relationships, we can express the mass of the neutron star in
terms of the distance to the system as:
$$
M_{\mathrm{PSR}} = \left(7.1\pm 0.1\right) \times
\left(\frac{D}{2.20\pm 0.02\ \mathrm{kpc}}\right)^{3} M_\Sun.
$$
Taking our distance measurement of $1368^{+42}_{-39}$~pc and adding
all three uncertainties in quadrature, we can obtain $M_{\mathrm{PSR}}
= 1.71 \pm 0.16 M_\Sun$, where the distance uncertainty remains the
dominant contributor to the error budget.  This result is consistent
with recent analyses which have showed that recycled pulsars have
masses of approximately $1.5 \pm 0.2 M_\Sun$ \citep{ozel12a,zhang11a}.

The assumption of Roche lobe filling by the companion of J1023 remains
a concern. While it seems likely that the active episode experienced by the
J1023 system in 2001 \citep[summarized in][]{archibald09a} was due
to Roche lobe overflow, we have regrettably limited information about
the nature of that episode. It is possible that the active episode was
actually the result of some other form of mass transfer, perhaps due
to very strong winds from the companion, though we deem this unlikely.
If the 2001 activity did indeed involve Roche lobe overflow,
\citet{thorstensen05a} point out that the Kelvin--Helmholtz relaxation time for the
companion is much longer than the several years since the active
episode, implying that the companion should still fill its Roche lobe.
Regardless, if the companion's radius is a fraction f (necessarily
less than one, or excess material would fall away) of the Roche lobe
radius, the pulsar mass is approximately $(1.71\pm 0.16 M_\Sun)\mathrm{f}^{-3}$. That
is, if the companion substantially underfills its Roche lobe, the mass
estimate we present is a lower limit for the pulsar mass.


It would be valuable to have observations that tested whether the
companion fills its Roche lobe. \citet{thorstensen05a} found that
models with a heated Roche--lobe--filling companion fit the optical
light curves very well, but they did not explicitly test models that
did not fill the Roche lobe. Sensitive optical photometric
observations might reveal more structure in the light curves, but
phase--resolved high--quality spectroscopy is an even more promising
avenue for probing the companion's size and shape.

\acknowledgements  The National Radio Astronomy Observatory is a facility of the National Science Foundation operated under cooperative agreement by Associated Universities, Inc.  The WSRT is operated by ASTRON with support from Netherlands Foundation for Scientific Research (NWO).  ATD was supported by an NRAO Jansky Fellowship and an NWO Veni Fellowship.  SC acknowledges support from the NSF through the award AST-1008213.  IHS received support from an NSERC Discovery Grant and the Canada Foundation for Innovation.  VMK holds the Lorne Trottier Chair in Astrophysics \& Cosmology, a Canada Research Chair, a Killam Research Fellowship, and acknowledges additional support from an NSERC Discovery Grant, from FQRNT via le Centre de Recherche Astrophysique du Qu\'ebec and the Canadian Institute for Advanced Research.  The authors thank Jason Hessels and Julia Deneva for their help performing timing observations with the Westerbork and Arecibo telescopes.

\bibliographystyle{apj}

\end{document}